\documentclass[11pt]{article}

\usepackage{a4wide}
\usepackage{multirow}
\usepackage{graphicx}
\usepackage{amsmath}
\usepackage{amssymb}
\usepackage{hyperref}

\def\beq{\begin{equation}}
\def\eeq{\end{equation}}
\def\be{\begin{equation}}
\def\ee{\end{equation}}

\newcommand\sss{\scriptscriptstyle\rm}
\newcommand\as{\alpha_{\sss S}}
\newcommand\muR{\mu_{\sss R}}
\newcommand\muF{\mu_{\sss F}}
\newcommand\xiR{\xi_{\sss R}}
\newcommand\xiF{\xi_{\sss F}}
\newcommand\xiRM{\xi_{\sss R}^{(max)}}
\newcommand\xiFM{\xi_{\sss F}^{(max)}}
\newcommand\xiRm{\xi_{\sss R}^{(min)}}
\newcommand\xiFm{\xi_{\sss F}^{(min)}}
\newcommand\xiRMb{\bar{\xi}_{\sss R}^{(max)}}
\newcommand\xiFMb{\bar{\xi}_{\sss F}^{(max)}}
\newcommand\xiRmb{\bar{\xi}_{\sss R}^{(min)}}
\newcommand\xiFmb{\bar{\xi}_{\sss F}^{(min)}}

\title{\textbf{Updated predictions for the total production cross
sections of top and of heavier quark pairs\\ at the Tevatron and at the LHC}}

\author{Matteo Cacciari$^a$, Stefano Frixione$^{b,c,}$\footnote{On leave 
of absence from INFN, Sezione di Genova, Genoa, Italy.},
Michelangelo L. Mangano$^b$,\\ 
Paolo Nason$^d$ and Giovanni Ridolfi$^{b,e}$\\[5pt]
\sl ~\\
$^a$ LPTHE, UPMC Universit\'e Paris 6,\\
Universit\'e Paris Diderot -- Paris 7, and CNRS UMR 7589, Paris, France\\[5pt]
$^b$ PH Department, TH Unit, CERN, CH-1211 Geneva 23, Switzerland\\[5pt]
$^c$ Institut de Th\'eorie des Ph\'enom\`enes Physiques, EPFL, 
CH-1015 Lausanne, Switzerland\\[5pt]
$^d$ INFN, Sezione di Milano-Bicocca,\\
Piazza della Scienza 3, 20126 Milan, Italy\\[5pt]
$^e$ Dipartimento di Fisica, Universit\`a di Genova, and INFN, 
Sezione di Genova,\\
Via Dodecaneso 33, I-16146 Genoa, Italy.\\[5pt]
E-mail: matteo.cacciari@lpthe.jussieu.fr, stefano.frixione@cern.ch, \\
michelangelo.mangano@cern.ch,
paolo.nason@mib.infn.it, giovanni.ridolfi@ge.infn.it
}

\date{}

\begin{document}

\maketitle

\vspace{3cm}
\begin{abstract}
We present updated predictions for the total production cross section
of top-quark pairs at the Tevatron and at the LHC, and, at the LHC,  
of heavy-quark pairs with mass in the range $0.5-2$~TeV.  
For $t\bar{t}$ production at the LHC we
also present results at $\sqrt{S}= 10$~TeV, in view of the expected accelerator
conditions during the forthcoming 2008 run. 
Our results are accurate at the
level of next-to-leading order in $\as$, and of next-to-leading
threshold logarithms (NLO+NLL). We adopt the most recent parametrizations
of parton distribution functions, and compute the corresponding
uncertainties. We study the dependence of the results on the top
mass, and we assess the impact of missing higher-order corrections
by independent variations of factorisation and renormalisation scales.
\end{abstract}

\vfill

%
\section{Top production\label{sec:top}}
One of the first tasks of the Large Hadron Collider (LHC) experiments
when, later in 2008, they will start taking data, will be to
re-discover the Standard Model. This will
serve the double purpose of making sure that the detectors are well
understood and work properly, as well as of improving the precision of
previous measurements. With a predicted cross section for top quark
pairs about a hundred
times larger than at the Fermilab Tevatron, and a much higher
design luminosity, the LHC is poised to become a real ``top
factory''~\cite{Beneke:2000hk}. 
This will allow for better measurements of the mass and the
cross section.  The former is expected to be measured with
an ultimate uncertainty below 1 GeV~\cite{Beneke:2000hk,Kharchilava:1999yj} 
(to be compared with the
most recent determination from the Tevatron, $m_t = 172.6 \pm
0.8\pm 1.1$~GeV~\cite{Group:2008nq}). 
The cross section is expected to be measured within a
year with a 15\% accuracy, and  eventually with an accuracy probably
limited only by the knowledge of the LHC luminosity, expected to reach
a precision of a few per cent~\cite{:2008tz}.

Experimental measurements of the total cross section at the
Tevatron~\cite{:2007qf,:2007bu} are usually compared to
predictions~\cite{Cacciari:2003fi,Kidonakis:2003qe} compiled a few
years ago\footnote{The precision of the cross section experimental
measurements has recently become sufficiently good that extractions of
the top mass by comparing the measured cross section with the
calculated value have become possible, and have been performed
\cite{:2007jw}.}. These predictions made use of the 
next-to-leading order (NLO) calculations of \cite{Nason:1987xz}, and
of the soft-gluon next-to-leading-log (NLL) threshold resummation
results obtained in \cite{Catani:1996yz,Bonciani:1998vc} and 
\cite{Kidonakis:2000ui} respectively. Some logarithmic
contributions of order higher than NLL have been included in the
results of ref.~\cite{Kidonakis:2003qe,Kidonakis:2001nj} and subsequent papers. 
In this way, while a complete NNLO calculation is still unavailable (the 
first ingredients,
two loop virtual corrections in the
ultrarelativistic limit \cite{Czakon:2007ej,Czakon:2007wk}, 
fully massive results for the two-loop contribution in the quark-quark 
channel~\cite{Czakon:2008zk,Bonciani:2008az} and for the
one-loop-squared contributions~\cite{Korner:2005rg,Korner:2008bn}, real emission
at one loop~\cite{Dittmaier:2007wz}, 
have been recently obtained), {\em some}
NNLO contributions of order $\as^4$,  of soft origin,  can be obtained
through an expansion and truncation of the Sudakov exponent. More
recently, while this paper was being completed,
the final ingredients required for a complete resummation of
soft-gluon next-to-next-to-leading logarithms (NNLL) have been
calculated~\cite{Moch:2008qy}, and the relative impact on the total
cross sections was explored.

A key feature of our most recent study in 
\cite{Cacciari:2003fi} was an extensive exploration of the theoretical
uncertainties affecting the prediction. The effect of the independent
variations of
renormalisation and of factorisation scales, which is the customary way
to assess the impact of unknown higher-order contributions, was
explored in detail. Parton distribution function sets (PDFs) providing 
a mean to
estimate the associated uncertainty \cite{Giele:2001mr} were also used. It was
determined in \cite{Cacciari:2003fi} that a significant fraction of
the overall $\pm 10-13\%$ uncertainty in $p\bar p$
production at the Tevatron was originating from the PDFs, though higher orders
were also contributing a fair share.
This result should now be revisited on a number of counts. First, new PDF sets
with errors, CTEQ6.5 \cite{Tung:2006tb}, MRST2006nnlo \cite{Martin:2007bv} and CTEQ 6.6
\cite{Nadolsky:2008zw} have appeared in the past few years.
It is legitimate to wonder if they might come with a reduced uncertainty.
Second, a similarly careful job of estimating the theoretical uncertainties
for the best available prediction should be made for the LHC too.  Third,
since the most recent Tevatron measurements point to a lower mass than the
central value $m_t=175$~GeV used in \cite{Cacciari:2003fi}, it is useful to
produce numerical predictions for an updated value of the top mass. 
Note that an analysis of the $t\bar{t}$ cross section has recently been
performed in \cite{Nadolsky:2008zw}. This study, carried out at the fixed-order, NLO level, 
focuses  on the correlations of the top cross section with other observables, 
analyzed  as a function of the PDF sets.

We shall present our results in the form
\begin{equation}
\sigma=
\sigma({\rm central})_{-\Delta{\sigma}_{\mu -}}^{+\Delta{\sigma}_{\mu +}}
~_{-\Delta{\sigma}_{\sss PDF-}}^{+\Delta{\sigma}_{\sss PDF+}}\,,
\end{equation}
where $\sigma({\rm central})$ is our best prediction, and 
$\Delta{\sigma}_{\mu\pm}$ and $\Delta{\sigma}_{\sss PDF\pm}$
quantify the uncertainties due to higher perturbative orders
and PDF choices, as specified in what follows.

In order to streamline the calculation of the overall uncertainty
(unknown higher orders and PDFs) we modify slightly the method
employed in ref. \cite{Cacciari:2003fi} and proceed as follows.

\begin{itemize}
\item Our best prediction $\sigma({\rm central})$ is computed by setting
the renormalisation and factorisation scales equal to $m_t$, and with
the central PDF set (within a given PDF error family). The cross section
is calculated to NLO+NLL accuracy, exactly as in ref.~\cite{Cacciari:2003fi}. 

\item The uncertainty on higher orders is estimated by varying the
factorisation and the renormalisation scales $\muF$ and $\muR$
independently around a central scale set by the top mass $m_t$. We
define the ratios 
\begin{equation}
\xiF=\muF/m_t\,,\;\;\;\;\;\;
\xiR=\muR/m_t\,,
\end{equation}
and we allow them to vary in the regions $0.5 \le \xiF,\xiR \le 2$, with
the condition that $0.5\le \xiF/\xiR \le 2$. 
This means that none of the ratios $\muF/m_t$,
$\muR/m_t$ and $\muF/\muR$ can be larger than two or
smaller than one-half, in order not to have in
the perturbative expansion logarithms of arguments larger than a
chosen (admittedly arbitrary) amount. Within this region the
NLO+NLL cross section is evaluated\footnote{A NLL resummation function 
with independent renormalisation and factorisation scales is given 
explicitly in~\cite{Cacciari:1999sy}.}, and used to compute\footnote{The
quantities $\Delta{\sigma}_{\mu +}$ and $\Delta{\sigma}_{\mu -}$ are
positive for all choices of top mass and scales we have considered.}
\begin{eqnarray}
&&\Delta{\sigma}_{\mu +} = \max_{\{\xiF,\xiR\}}
\Big[\sigma(\xiF,\xiR)-\sigma(1,1)\Big]\,,
\label{eq:muplus}
\\
&&\Delta{\sigma}_{\mu -} = -\min_{\{\xiF,\xiR\}}
\Big[\sigma(\xiF,\xiR)-\sigma(1,1)\Big]\,.
\label{eq:muminus}
\end{eqnarray}
All cross sections in these formulae are evaluated with the
central PDF set (thus, $\sigma(1,1)\equiv\sigma({\rm central})$ here).
We also introduce the symbols
\begin{eqnarray}
{\rm Scales}+&=&\sigma({\rm central})+\Delta{\sigma}_{\mu +}
\label{eq:defScpl}
\\*
{\rm Scales}-&=&\sigma({\rm central})-\Delta{\sigma}_{\mu -}
\label{eq:defScmn}
\end{eqnarray}
which we shall use in the following.

By doing so we have established a variation interval  of the cross
section that can be considered as a reasonable estimate of the
uncertainty due to unknown higher orders. It should be noted,
however, that such an uncertainty should by no means be considered as
distributed according to some probability law 
(for instance, with a Gaussian distribution) around the
central value. In fact, it is  more similar to a systematic than to a
statistical uncertainty. This means that further arbitrary choices will
have to be made in order to assign a `confidence level' to this
interval. 

\item Modern PDF sets come with a procedure to evaluate the propagation
of their uncertainty onto a given physical observable. This is done by
exploring the effect of using, along with a `central' PDF set, a number of
other sets (usually 40 for the CTEQ family PDFs, 30 for the MRST family
ones) and properly combining their differences. According to
the CTEQ and MRST Collaborations, the resulting uncertainty should roughly
represent a 90\% confidence level. We have chosen to follow the prescription 
by Nadolsky and Sullivan~\cite{Nadolsky:2001yg}, and determine asymmetric 
uncertainties in the form
\begin{eqnarray}
&&\Delta{\sigma}_{\sss PDF+} = 
\sqrt{\sum_i\Big(\max\Big[{\sigma}(set_{+i}) - {\sigma}(set_0),
 {\sigma}(set_{-i}) - {\sigma}(set_0), 0\Big]\Big)^2} \,, 
\label{eq:PDFplus}
\\
&&\Delta{\sigma}_{\sss PDF-} = 
\sqrt{\sum_i\Big(\max\Big[{\sigma}(set_0) - {\sigma}(set_{+i}),
 {\sigma}(set_0) - {\sigma}(set_{-i}), 0\Big]\Big)^2} \,.
\label{eq:PDFminus}
\end{eqnarray}
where all cross sections are evaluated with
\begin{equation}
\xiF=1\,,\;\;\;\;\;\;\xiR=1\,.
\end{equation}
In eqs.~(\ref{eq:PDFplus}) and~(\ref{eq:PDFminus}), $set_0$ represents 
the central set, and the sums run over all pairs of PDFs in the given 
PDF error set. For each pair, we denote by $set_{+i}$ and $set_{-i}$ 
the positive and negative displacement member of the pair. 
We also introduce the symbols
\begin{eqnarray}
{\rm PDFs}+&=&\sigma({\rm central})+\Delta{\sigma}_{\sss PDF+}
\label{eq:defmupl}
\\*
{\rm PDFs}-&=&\sigma({\rm central})-\Delta{\sigma}_{\sss PDF-}
\label{eq:defmumn}
\end{eqnarray}
which we shall use in the following.
\end{itemize}
To facilitate the determination of theoretical cross section
corresponding to mass values different than the current best fit, we
provide our results in the form of the
coefficients of the parametrization
\begin{equation}
\sigma(m_t) = A + B (m_t-171) + 
C (m_t-171)^2 + D (m_t - 171)^3 \; .
\end{equation}
The parameters were fitted to the exact results in the range $150 \le m_t \le
190$~GeV, with a precision of the order of 1-2 per mille. The $A$
coefficient has been fixed equal to the cross section at $m_t =
171$~GeV.
The fit parameters are
 given in table \ref{table:tevatron} and table  \ref{table:lhc}, for
the Tevatron and the LHC respectively\footnote{Since the LHC is
  scheduled
to run in 2008 at a
centre of mass energy of 10 TeV, we have also provided predictions for
this energy in table \ref{table:lhc10}, in the case a measurement of the
total cross section of top production should prove possible.}.

For each PDF set we have listed separately the `central' value  (scales $=
m_t$, central PDF set) and the maximum and the minimum found by varying the
scales  according to the above procedure and evaluating the asymmetric PDFs
uncertainties. 
The main effects of the different PDFs and of the uncertainties can of course
be read off directly from the $A$ coefficient, which corresponds to the
$t\bar t$ cross section evaluated at $m_t = 171$~GeV. The display of
results obtained with many
different PDF sets, both very recent and older, is meant to allow for an
easy estimate of the variation (or lack thereof) of the cross section
predictions as
a consequence of evolving parton distribution functions sets.

\begin{table}
\begin{center}
\begin{tabular}{|c|l|c|c|c|c|}
\hline
\multicolumn{2}{|c|}{Tevatron, $p\bar p$ at $\sqrt{s}=1960$ GeV } & $A$ (pb) & $B$ (pb/GeV) & $C$ (pb/GeV$^2$) & $D$ (pb/GeV$^3$) \\
\hline
\multirow{5}{*}{CTEQ6M}
& Central & 7.59 & -0.237 & 4.39 $\times 10^{-3}$ & -6.32 $\times 10^{-5}$ \\ 
& Scales$+$ & 7.89 & -0.247 & 4.60 $\times 10^{-3}$ & -6.66 $\times 10^{-5}$ \\ 
& Scales$-$ & 7.07 & -0.221 & 4.11 $\times 10^{-3}$ & -5.92 $\times 10^{-5}$ \\ 
& PDFs$+$ & 8.26 & -0.260 & 4.86 $\times 10^{-3}$ & -7.02 $\times 10^{-5}$ \\ 
& PDFs$-$ & 7.12 & -0.222 & 4.08 $\times 10^{-3}$ & -5.82 $\times 10^{-5}$ \\ 
\hline
\multirow{5}{*}{CTEQ6.1}
& Central & 7.77 & -0.244 & 4.53 $\times 10^{-3}$ & -6.51 $\times 10^{-5}$ \\ 
& Scales$+$ & 8.08 & -0.254 & 4.74 $\times 10^{-3}$ & -6.86 $\times 10^{-5}$ \\
& Scales$-$ & 7.23 & -0.227 & 4.23 $\times 10^{-3}$ & -6.09 $\times 10^{-5}$ \\
& PDFs$+$ & 8.53 & -0.269 & 5.04 $\times 10^{-3}$ & -7.27 $\times 10^{-5}$ \\ 
& PDFs$-$ & 7.20 & -0.224 & 4.12 $\times 10^{-3}$ & -5.87 $\times 10^{-5}$ \\ 
\hline
\multirow{5}{*}{CTEQ6.5}
& Central & 7.61 & -0.237 & 4.38 $\times 10^{-3}$ & -6.28 $\times 10^{-5}$ \\ 
& Scales$+$ & 7.90 & -0.247 & 4.58 $\times 10^{-3}$ & -6.61 $\times 10^{-5}$ \\ 
& Scales$-$ & 7.08 & -0.221 & 4.10 $\times 10^{-3}$ & -5.89 $\times 10^{-5}$ \\ 
& PDFs$+$ & 8.14 & -0.256 & 4.78 $\times 10^{-3}$ & -6.91 $\times 10^{-5}$ \\ 
& PDFs$-$ & 7.24 & -0.224 & 4.11 $\times 10^{-3}$ & -5.85 $\times 10^{-5}$ \\ 
\hline
\multirow{5}{*}{CTEQ6.6}
& Central & 7.48 & -0.233 & 4.32 $\times 10^{-3}$ & -6.20 $\times 10^{-5}$ \\ 
& Scales $+$ & 7.77 & -0.243 & 4.52 $\times 10^{-3}$ & -6.53 $\times 10^{-5}$ \\ 
& Scales $-$ & 6.96 & -0.218 & 4.04 $\times 10^{-3}$ & -5.80 $\times 10^{-5}$ \\ 
& PDFs $+$ & 7.99 & -0.251 & 4.70 $\times 10^{-3}$ & -6.79 $\times 10^{-5}$ \\ 
& PDFs $-$ & 7.09 & -0.220 & 4.02 $\times 10^{-3}$ & -5.72 $\times 10^{-5}$ \\ 
\hline
\multirow{5}{*}{MRST2001E}
& Central & 7.66 & -0.242 & 4.53 $\times 10^{-3}$ & -6.60 $\times 10^{-5}$ \\ 
& Scales$+$ & 7.97 & -0.252 & 4.75 $\times 10^{-3}$ & -6.98 $\times 10^{-5}$ \\ 
& Scales$-$ & 7.13 & -0.225 & 4.24 $\times 10^{-3}$ & -6.17 $\times 10^{-5}$ \\ 
& PDFs$+$ & 7.94 & -0.252 & 4.75 $\times 10^{-3}$ & -6.95 $\times 10^{-5}$ \\ 
& PDFs$-$ & 7.44 & -0.233 & 4.35 $\times 10^{-3}$ & -6.31 $\times 10^{-5}$ \\ 
\hline
MRST2004nlo
& Central & 7.99 & -0.253 & 4.77 $\times 10^{-3}$ & -6.95 $\times 10^{-5}$ \\ 
\hline
\multirow{5}{*}{MRST2006nnlo}
& Central & 7.93 & -0.253 & 4.76 $\times 10^{-3}$ & -6.92 $\times 10^{-5}$ \\ 
& Scales$+$ & 8.27 & -0.264 & 5.00 $\times 10^{-3}$ & -7.33 $\times 10^{-5}$ \\ 
& Scales$-$ & 7.37 & -0.235 & 4.44 $\times 10^{-3}$ & -6.45 $\times 10^{-5}$ \\ 
& PDFs$+$ & 8.17 & -0.261 & 4.93 $\times 10^{-3}$ & -7.19 $\times 10^{-5}$ \\ 
& PDFs$-$ & 7.73 & -0.245 & 4.61 $\times 10^{-3}$ & -6.68 $\times 10^{-5}$ \\ 
\hline
\end{tabular}
\end{center}
\caption{\label{table:tevatron}
Coefficients of the parametrization $\sigma(m_t) = A + B (m_t-171) + C
(m_t-171)^2 + D (m_t - 171)^3$} for the NLO+NLL $t\bar t$ cross section 
(picobarn) at the Tevatron, for various PDF sets. The fit must not be
used outside the range $150 \le m_t \le 190$~GeV. The quantities
Scales$\pm$ and PDFs$\pm$ are defined in eqs.~(\ref{eq:defScpl}),
(\ref{eq:defScmn}), (\ref{eq:defmupl}), and (\ref{eq:defmumn}).
\end{table}

\begin{table}
\begin{center}
\begin{tabular}{|c|l|c|c|c|c|}
\hline
\multicolumn{2}{|c|}{LHC, $pp$ at $\sqrt{s}=10$ TeV } & $A$ (pb) & $B$ (pb/GeV) & $C$ (pb/GeV$^2$) & $D$ (pb/GeV$^3$) \\
\hline
\multirow{5}{*}{CTEQ6M}
& Central & 425 & -12.1 & 0.211 & -2.89 $\times 10^{-3}$ \\ 
& Scales$+$ & 462 & -13.2 & 0.232 & -3.20 $\times 10^{-3}$ \\ 
& Scales$-$ & 386 & -10.9 & 0.189 & -2.58 $\times 10^{-3}$ \\ 
& PDFs$+$ & 445 & -12.5 & 0.216 & -2.94 $\times 10^{-3}$ \\ 
& PDFs$-$ & 406 & -11.7 & 0.205 & -2.82 $\times 10^{-3}$ \\ 
\hline
\multirow{5}{*}{CTEQ6.1}
& Central & 428 & -12.1 & 0.211 & -2.87 $\times 10^{-3}$ \\ 
& Scales$+$ & 465 & -13.2 & 0.232 & -3.19 $\times 10^{-3}$ \\ 
& Scales$-$ & 389 & -10.9 & 0.189 & -2.57 $\times 10^{-3}$ \\ 
& PDFs$+$ & 450 & -12.5 & 0.216 & -2.93 $\times 10^{-3}$ \\ 
& PDFs$-$ & 406 & -11.6 & 0.205 & -2.81 $\times 10^{-3}$ \\ 
\hline
\multirow{5}{*}{CTEQ6.5}
& Central & 414 & -11.7 & 0.205 & -2.79 $\times 10^{-3}$ \\ 
& Scales$+$ & 450 & -12.9 & 0.226 & -3.09 $\times 10^{-3}$ \\ 
& Scales$-$ & 376 & -10.6 & 0.184 & -2.50 $\times 10^{-3}$ \\ 
& PDFs$+$ & 434 & -12.2 & 0.211 & -2.85 $\times 10^{-3}$ \\ 
& PDFs$-$ & 396 & -11.3 & 0.199 & -2.72 $\times 10^{-3}$ \\ 
\hline
\multirow{5}{*}{CTEQ6.6}
& Central & 414 & -11.8 & 0.206 & -2.81 $\times 10^{-3}$ \\ 
& Scales $+$ & 451 & -12.9 & 0.227 & -3.12 $\times 10^{-3}$ \\ 
& Scales $-$ & 376 & -10.6 & 0.185 & -2.51 $\times 10^{-3}$ \\ 
& PDFs $+$ & 433 & -12.2 & 0.211 & -2.86 $\times 10^{-3}$ \\ 
& PDFs $-$ & 396 & -11.4 & 0.200 & -2.75 $\times 10^{-3}$ \\ 
\hline
\multirow{5}{*}{MRST2001E}
& Central & 446 & -12.6 & 0.217 & -2.94 $\times 10^{-3}$ \\ 
& Scales$+$ & 486 & -13.8 & 0.240 & -3.27 $\times 10^{-3}$ \\ 
& Scales$-$ & 405 & -11.3 & 0.195 & -2.63 $\times 10^{-3}$ \\ 
& PDFs$+$ & 457 & -12.8 & 0.220 & -2.97 $\times 10^{-3}$ \\ 
& PDFs$-$ & 439 & -12.4 & 0.216 & -2.92 $\times 10^{-3}$ \\ 
\hline
MRST2004nlo
& Central & 455 & -12.8 & 0.221 & -2.99 $\times 10^{-3}$ \\ 
\hline
\multirow{5}{*}{MRST2006nnlo}
& Central & 446 & -12.5 & 0.216 & -2.92 $\times 10^{-3}$ \\ 
& Scales$+$ & 486 & -13.7 & 0.238 & -3.24 $\times 10^{-3}$ \\ 
& Scales$-$ & 404 & -11.3 & 0.194 & -2.60 $\times 10^{-3}$ \\ 
& PDFs$+$ & 454 & -12.7 & 0.218 & -2.93 $\times 10^{-3}$ \\ 
& PDFs$-$ & 438 & -12.3 & 0.214 & -2.89 $\times 10^{-3}$ \\ 
\hline
\end{tabular}
\end{center}
\caption{\label{table:lhc10}
Coefficients of the parametrization $\sigma(m_t) = A + B (m_t-171) + C
(m_t-171)^2 + D (m_t - 171)^3$} for the  NLO+NLL $t\bar t$ cross section 
(picobarn) at the LHC with $\sqrt{s} = 10$~TeV, for various PDF sets.
The fit must not be used outside the range $150 \le m_t \le 190$~GeV.
The quantities 
Scales$\pm$ and PDFs$\pm$ are defined in eqs.~(\ref{eq:defScpl}),
(\ref{eq:defScmn}), (\ref{eq:defmupl}), and (\ref{eq:defmumn}).
\end{table}

\begin{table}
\begin{center}
\begin{tabular}{|c|l|c|c|c|c|}
\hline
\multicolumn{2}{|c|}{LHC, $pp$ at $\sqrt{s}=14$ TeV } & $A$ (pb) & $B$ (pb/GeV) & $C$ (pb/GeV$^2$) & $D$ (pb/GeV$^3$) \\
\hline
\multirow{5}{*}{CTEQ6M}
& Central & 933 & -25.3 & 0.423 & -5.60 $\times 10^{-3}$ \\ 
& Scales$+$ & 1018 & -27.7 & 0.468 & -6.22 $\times 10^{-3}$ \\ 
& Scales$-$ & 846 & -22.8 & 0.379 & -4.99 $\times 10^{-3}$ \\ 
& PDFs$+$ & 962 & -25.8 & 0.432 & -5.73 $\times 10^{-3}$ \\ 
& PDFs$-$ & 903 & -24.6 & 0.413 & -5.44 $\times 10^{-3}$ \\ 
\hline
\multirow{5}{*}{CTEQ6.1}
& Central & 934 & -25.2 & 0.421 & -5.56 $\times 10^{-3}$ \\ 
& Scales$+$ & 1019 & -27.7 & 0.466 & -6.19 $\times 10^{-3}$ \\ 
& Scales$-$ & 847 & -22.7 & 0.377 & -4.95 $\times 10^{-3}$ \\ 
& PDFs$+$ & 965 & -25.8 & 0.430 & -5.70 $\times 10^{-3}$ \\ 
& PDFs$-$ & 902 & -24.5 & 0.411 & -5.40 $\times 10^{-3}$ \\ 
\hline
\multirow{5}{*}{CTEQ6.5}
& Central & 908 & -24.5 & 0.411 & -5.46 $\times 10^{-3}$ \\ 
& Scales$+$ & 990 & -26.9 & 0.455 & -6.08 $\times 10^{-3}$ \\ 
& Scales$-$ & 823 & -22.1 & 0.368 & -4.87 $\times 10^{-3}$ \\ 
& PDFs$+$ & 938 & -25.2 & 0.420 & -5.57 $\times 10^{-3}$ \\ 
& PDFs$-$ & 879 & -23.9 & 0.401 & -5.29 $\times 10^{-3}$ \\ 
\hline
\multirow{5}{*}{CTEQ6.6}
& Central & 911 & -24.7 & 0.413 & -5.47 $\times 10^{-3}$ \\ 
& Scales $+$ & 993 & -27.1 & 0.457 & -6.09 $\times 10^{-3}$ \\ 
& Scales $-$ & 826 & -22.2 & 0.370 & -4.87 $\times 10^{-3}$ \\ 
& PDFs $+$ & 939 & -25.2 & 0.422 & -5.58 $\times 10^{-3}$ \\ 
& PDFs $-$ & 881 & -24.0 & 0.404 & -5.36 $\times 10^{-3}$ \\ 
\hline
\multirow{5}{*}{MRST2001E}
& Central & 965 & -25.9 & 0.429 & -5.63 $\times 10^{-3}$ \\ 
& Scales$+$ & 1054 & -28.4 & 0.475 & -6.27 $\times 10^{-3}$ \\ 
& Scales$-$ & 874 & -23.3 & 0.384 & -5.00 $\times 10^{-3}$ \\ 
& PDFs$+$ & 981 & -26.2 & 0.434 & -5.68 $\times 10^{-3}$ \\ 
& PDFs$-$ & 954 & -25.6 & 0.426 & -5.57 $\times 10^{-3}$ \\ 
\hline
MRST2004nlo
& Central & 982 & -26.3 & 0.436 & -5.72 $\times 10^{-3}$ \\ 
\hline
\multirow{5}{*}{MRST2006nnlo}
& Central & 961 & -25.7 & 0.426 & -5.58 $\times 10^{-3}$ \\ 
& Scales$+$ & 1050 & -28.3 & 0.472 & -6.21 $\times 10^{-3}$ \\ 
& Scales$-$ & 870 & -23.1 & 0.381 & -4.96 $\times 10^{-3}$ \\ 
& PDFs$+$ & 972 & -25.9 & 0.428 & -5.62 $\times 10^{-3}$ \\ 
& PDFs$-$ & 949 & -25.4 & 0.422 & -5.53 $\times 10^{-3}$ \\ 
\hline
\end{tabular}
\end{center}
\caption{\label{table:lhc}
Coefficients of the parametrization $\sigma(m_t) = A + B (m_t-171) + C
(m_t-171)^2 + D (m_t - 171)^3$} for the  NLO+NLL $t\bar t$ cross section 
(picobarn) at the LHC, for various PDF sets. The fit must not be
used outside the range $150 \le m_t \le 190$~GeV. The quantities 
Scales$\pm$ and PDFs$\pm$ are defined in eqs.~(\ref{eq:defScpl}),
(\ref{eq:defScmn}), (\ref{eq:defmupl}), and (\ref{eq:defmumn}).
\end{table}

We summarise here what might be considered our ``best'' predictions for
$t\bar t$ production at the LHC, at $m_t = 171$~GeV:
\begin{equation}
\sigma^\mathrm{NLO+NLL}_{t\bar t}(\mathrm{LHC}, m_t = 171~\mathrm{GeV},
\mathrm{CTEQ6.5})
= 908 {~}^{+82(9.0\%)}_{-85(9.3\%)}~\mathrm{(scales)} 
{~}^{+30(3.3\%)}_{-29(3.2\%)} ~\mathrm{(PDFs)} ~~\mathrm{pb}
\end{equation}
\begin{equation}
\sigma^\mathrm{NLO+NLL}_{t\bar t}(\mathrm{LHC}, m_t = 171~\mathrm{GeV},
\mathrm{MRST2006nnlo})
= 961 {~}^{+89(9.2\%)}_{-91(9.4\%)}~\mathrm{(scales)} 
{~}^{+11(1.1\%)}_{-12(1.2\%)} ~\mathrm{(PDFs)} ~~\mathrm{pb}
\end{equation}
Note that we quote separately the
results for MRST and CTEQ, since they are not fully consistent. 
For reference, we also give here the pure fixed-order results (i.e.,
without including threshold resummation) at the NLO and the LO
\begin{equation}
\sigma^\mathrm{NLO}_{t\bar t}(\mathrm{LHC}, m_t = 171~\mathrm{GeV},
\mathrm{CTEQ6.5})
= 875 {~}^{+102(11.6\%)}_{-100(11.5\%)}~\mathrm{(scales)} 
{~}^{+30(3.4\%)}_{-29(3.3\%)} ~\mathrm{(PDFs)} ~~\mathrm{pb}
\label{eq:ttLHCcteqNLO}
\end{equation}
\begin{equation}
\sigma^\mathrm{LO}_{t\bar t}(\mathrm{LHC}, m_t = 171~\mathrm{GeV},
\mathrm{CTEQ6.5})
= 583 {~}^{+165(28.2\%)}_{-120(20.7\%)}~\mathrm{(scales)} 
{~}^{+20(3.4\%)}_{-19(3.3\%)} ~\mathrm{(PDFs)} ~~\mathrm{pb}
\end{equation}

\begin{equation}
\sigma^\mathrm{NLO}_{t\bar t}(\mathrm{LHC}, m_t = 171~\mathrm{GeV},
\mathrm{MRST2006nnlo})
= 927 {~}^{+109(11.7\%)}_{-107(11.5\%)}~\mathrm{(scales)} 
{~}^{+11(1.2\%)}_{-12(1.3\%)} ~\mathrm{(PDFs)} ~~\mathrm{pb}
\end{equation}
\begin{equation}
\sigma^\mathrm{LO}_{t\bar t}(\mathrm{LHC}, m_t = 171~\mathrm{GeV},
\mathrm{MRST2006nnlo})
= 616 {~}^{+172(27.9\%)}_{-126(20.5\%)}~\mathrm{(scales)} 
{~}^{+7.3(1.2\%)}_{-7.8(1.3\%)} ~\mathrm{(PDFs)} ~~\mathrm{pb}
\label{eq:ttLHCMRSTLO}
\end{equation}
We have decided not to combine the scales and PDFs uncertainties into
a single error. The reason for refraining from doing so is that the
scales uncertainty (and, to some extent, probably also the PDFs one) is
not fully characterised in statistical terms. 
As a consequence,
additional hypotheses will be needed in order to combine the two
uncertainties into a single probability density function for the
resulting cross section, with well defined confidence levels. 
Further discussions on the interplay between scales and PDFs uncertainties 
can be found in Appendix ~\ref{sec:details}.

We finally present our ``best'' predictions for
$t\bar t$ production at the Tevatron, at $m_t = 171$~GeV:
\begin{equation}
\sigma^\mathrm{NLO+NLL}_{t\bar t}(\mathrm{Tev}, m_t = 171~\mathrm{GeV},
\mathrm{CTEQ6.5})
= 7.61 {~}^{+0.30(3.9\%)}_{-0.53(6.9\%)}~\mathrm{(scales)} 
{~}^{+0.53(7\%)}_{-0.36(4.8\%)} ~\mathrm{(PDFs)} ~~\mathrm{pb}
\end{equation}
\begin{equation}
\sigma^\mathrm{NLO+NLL}_{t\bar t}(\mathrm{Tev}, m_t = 171~\mathrm{GeV},
\mathrm{MRST2006nnlo})
= 7.93 {~}^{+0.34(4.3\%)}_{-0.56(7.1\%)}~\mathrm{(scales)} 
{~}^{+0.24(3.1\%)}_{-0.20(2.5\%)} ~\mathrm{(PDFs)} ~~\mathrm{pb}\,.
\end{equation}
As done for the LHC in eqs.~(\ref{eq:ttLHCcteqNLO})--(\ref{eq:ttLHCMRSTLO}),
we also report the NLO and LO results:
\begin{equation}
\sigma^\mathrm{NLO}_{t\bar t}(\mathrm{Tev}, m_t = 171~\mathrm{GeV},
\mathrm{CTEQ6.5})
= 7.35 {~}^{+0.38(5.1\%)}_{-0.80(10.9\%)}~\mathrm{(scales)} 
{~}^{+0.49(6.6\%)}_{-0.34(4.6\%)} ~\mathrm{(PDFs)} ~~\mathrm{pb}
\end{equation}
\begin{equation}
\sigma^\mathrm{LO}_{t\bar t}(\mathrm{Tev}, m_t = 171~\mathrm{GeV},
\mathrm{CTEQ6.5})
= 5.92 {~}^{+2.34(39.5\%)}_{-1.54(26.1\%)}~\mathrm{(scales)} 
{~}^{+0.32(5.5\%)}_{-0.24(4.1\%)} ~\mathrm{(PDFs)} ~~\mathrm{pb}
\end{equation}

\begin{equation}
\sigma^\mathrm{NLO}_{t\bar t}(\mathrm{Tev}, m_t = 171~\mathrm{GeV},
\mathrm{MRST2006nnlo})
= 7.62 {~}^{+0.45(5.9\%)}_{-0.88(11.6\%)}~\mathrm{(scales)} 
{~}^{+0.23(3\%)}_{-0.18(2.4\%)} ~\mathrm{(PDFs)} ~~\mathrm{pb}
\end{equation}
\begin{equation}
\sigma^\mathrm{LO}_{t\bar t}(\mathrm{Tev}, m_t = 171~\mathrm{GeV},
\mathrm{MRST2006nnlo})
= 6.05 {~}^{+2.47(40.8\%)}_{-1.61(26.6\%)}~\mathrm{(scales)} 
{~}^{+0.16(2.6\%)}_{-0.13(2.1\%)} ~\mathrm{(PDFs)} ~~\mathrm{pb}\,.
\end{equation}

\section{Discussion of the $t\bar{t}$ cross section results at the LHC}
\label{sec:disc}
For all parameter choices we have considered, the scales uncertainties 
affecting the $t\bar t$ cross section at the LHC are much larger than 
those due to PDFs. In this section, we therefore focus on exploring 
the effect of the NLL resummation on the cross section. 
We can do so by comparing the
NLO prediction at $m_t = 171$ GeV, and its uncertainty due to scales
variations as described above, in the NLO and NLO+NLL approximations
respectively. We find that the `central' prediction is increased by
less than 4\%, and the scales uncertainty is only very mildly affected,
going from $\pm 11.5\%$ in the NLO case to $\pm 9\%$ in the NLO+NLL
one. This points to a relatively minor impact of threshold resummation on the
LHC cross section, as expected as a consequence of the relatively large
distance of the $t\bar t$ production threshold from the LHC centre of
mass energy (for comparison, the uncertainty at the Tevatron is almost
halved when going from NLO to NLO+NLL). One should also note that, again
contrary to the Tevatron case, exploring {\sl independent} scale
variations has a non-negligible effect: keeping $\muR = \muF$ 
(as done e.g. in ref.~\cite{Nadolsky:2008zw}) would
result in an uncertainty estimate for the NLO+NLL case of only
$^{+7}_{-5}\%$. We remind the reader that the fact PDF
fits are performed with $\muR=\muF$ does not force
us to use $\muR=\muF$ in the cross section. In fact
an independent variation of the two scales in our matched calculation
leads to variations in the result that are beyond the NLO+NLL
approximation. It is thus legitimate to add this independent variation
to the sources of uncertainties. It then turns out that the $\muR \neq
\muF$ approach leads to a much larger variation. We thus conclude that
there may be accidental cancellation in the scale variation when one
keeps $\muR=\muF$, leading to an unreliably small estimate of the error.

Another important element in the assessment of the systematics related
to the resummation is the estimate of the impact of beyond-NLL
corrections. To parametrize these corrections, a constant $A$ was introduced
  in~\cite{Bonciani:1998vc} 
  (where more details about its role are
  given):
\begin{equation}
\label{eq:Acoeff}
C_{ij} \to C_{ij} \; \left( 1-\frac{A}{N+A-1} \right), \quad
ij=q\bar{q},gg\; .
\end{equation} 
$C_{ij}$ here represents the $N$-independent term of the Mellin
transform of the NLO partonic cross section. The replacement in the
previous equation gives vanishing first moments, is irrelevant for large $N$,
and does not introduce poles on the real $N$ axis. Different choices of
$A$ give rise to different resummed cross sections, all consistent
with each other at the NLL level and NNLL level\footnote{Changing
$A$ corresponds to vary terms suppressed by powers of $N$, which
cannot be determined within the soft gluon approximation.}. 
They therefore parametrize the
possible exponentiation of finite, non-logarithmic terms appearing at
orders higher than NLO. It was noticed already
in~\cite{Bonciani:1998vc} that the choice $A=0$ was leading to a scale
dependence typically a factor of two smaller than was obtained with
$A=2$. For the sake of being conservative, we therefore selected $A=2$
in our subsequent phenomenological analysis~\cite{Cacciari:2003fi}, as
well as in the results presented in the previous section. We would
like to reiterate here this observation, by showing the results that
we would have obtained at the LHC if we had chosen $A=0$. In the case
of the CTEQ6.5 PDF, and $m_t=171$~GeV, the scale dependence obtained
by varying $\muF$ and $\muR$ independently is:
\begin{equation}
\sigma^\mathrm{NLO+NLL(A=0)}_{t\bar t}(\mathrm{LHC}, 
m_t = 171~\mathrm{GeV}, \mathrm{CTEQ6.5})
= 945 {~}^{+95(10\%)}_{-85(9.0\%)}~\mathrm{(scales)}~~\mathrm{pb} \qquad [\muF \neq \muR]\; ,
\end{equation}
which is approximately 5\% larger than the value obtained with $A=2$, a
  variation consistent with the estimated uncertainty. 
If we had chosen to set $\muF=\muR$, the result would have been:
\begin{equation}
\sigma^\mathrm{NLO+NLL(A=0)}_{t\bar t}(\mathrm{LHC}, 
m_t = 171~\mathrm{GeV}, \mathrm{CTEQ6.5})
= 945 {~}^{+19(2\%)}_{-7(0.7\%)}~\mathrm{(scales)}~~\mathrm{pb}\qquad [\muF = \muR] \; .
\end{equation}
Notice that the combination of $A=0$ and $\muR=\muF$ leads to a
dramatic reduction of the uncertainty. In particular, the reduction is
significant also with respect to the $A=2$, $\muR=\muF$ case,
discussed above. The choice $A=0$ and $\muR=\muF$ is
what was used in the recent analysis of the resummed NNLL
cross section of~\cite{Moch:2008qy}, leading to a similar uncertainty,
at the 2\% level. We conclude that it is not as yet clear whether the
improvement found in~\cite{Moch:2008qy} is a genuine reduction of the
uncertainty, that would survive the independent variation of 
$\muR$ and $\muF$ and with the introduction of a parameter similar to
our $A$.

We also note that the resummed NLO+NLL results quoted
in ref.~\cite{Moch:2008qy}  differ from ours to the extent of a few percent.
We have checked that this is not due to the choice made in \cite{Moch:2008qy} 
of limiting the contribution of NLL resummation to a $t\bar{t}$ mass range near
the production threshold, and using only the NLO result above such range 
(contrary to what stated in \cite{Moch:2008qy}, we do not perform the matching in
this way, but follow instead the procedure detailed in \cite{Bonciani:1998vc}). 
Rather, the small discrepancy is due to the misleading way in which 
eq.~(\ref{eq:Acoeff}) is presented in~\cite{Bonciani:1998vc}, and in which it 
was accordingly interpreted and used in~\cite{Moch:2008qy}.  There is in fact 
a mismatch in the Mellin $N$ argument appearing in the expression of coefficient
 $C_{ij}$ used in eqs. (54) and (58) of~\cite{Bonciani:1998vc}, with $N+1$ being
  the correct argument of the resummation function, rather than $N$. The 
  numerical implementation of these relations, in~\cite{Bonciani:1998vc} and 
  in this paper, are nevertheless consistent, and equivalent to a shift 
  $N\to N+1$ in~(\ref{eq:Acoeff}).
 An erratum to clarify this issue has been 
submitted for~\cite{Bonciani:1998vc}.

Another issue we wish to comment on is the relative size of the
scale and the PDFs uncertainty. It was observed in \cite{Cacciari:2003fi} that
the latter was important, and almost dominant, at the Tevatron. This
appears not to be the case anymore at the LHC: again according to the
procedure described above, at $m_t = 171$ GeV we find uncertainties of
the order of $\pm 3 \%$ with CTEQ6.5 and  $\pm 1.5 \%$ with MRST2006nnlo
and MRST2001E. The main reason for this improvement is that at the LHC
the range of $x$ values for the partons relevant to top production is
much smaller than at the Tevatron, and falls in a region where the
experimental knowledge of both quark and gluon PDFs is much better
constrained by data.
It is worth noting that the central results given by 
these two PDF sets, differing by about $6\%$, are not fully 
compatible, despite (or because of) the apparently very small estimated 
uncertainty. We also point out the MRST and CTEQ use different conventions
for the Tolerance parameter; the consequence of this is that, had the two
collaborations followed exactly the same fitting procedure, the PDF
uncertainty resulting from using an MRST family set would still be 
a factor of about $\sqrt{2}$ smaller than that obtained with a CTEQ set.

While the PDFs uncertainty is probably still somewhat underestimated,
as shown by the partially conflicting central values of
CTEQ6.5 and MRST2006nnlo, it is probably safe to conclude that, the 
very interesting progress with the NNLL resummation~\cite{Moch:2008qy}
notwithstanding, 
a definitive assessment of
our understanding of the $t\bar t$ cross section at
the LHC will have to wait for the full, massive NNLO calculation.

\section{Very heavy fermion production}
We now present production rates for a pair of fermions (belonging
to the fundamental representation of $SU(3)_{colour}$) heavier than top, 
using the same computations described for the $t\bar{t}$ cross
sections.  We shall 
generically denote such fermion pair by $T\bar{T}$. These particles
arise naturally in BSM theories with strongly-coupled dynamics; they
can be of different species, which can for example be classified
according to their transformation properties under 
SU(2)$_{\rm L}\otimes$SU(2)$_{\rm R}\otimes$U(1).
As far as pair production is concerned, however, these details are
largely irrelevant, since this process is expected to be dominated 
by QCD effects, and this is the reason why we can apply our NLO+NLL,
NLO, and LO results to the computations of $T\bar{T}$ rates.
In doing so, we shall neglect possible contributions of non-SM intermediate 
states resulting from e.g. a $q\bar{q}$ annihilation. Our aim is therefore 
not that of providing a complete phenomenological study of $T\bar{T}$ 
cross sections at the LHC, but rather that of assessing the scale and PDF
uncertainties affecting the QCD contribution to the
production of heavy fermion pairs.
In what follow, we shall consider the mass range 0.5~TeV$\le m_T\le$~2~TeV
for the heavy fermion. 

\begin{table}
\begin{center}
\begin{tabular}{|c|c|c|c|}
\hline
$m_T$ & NLO+NLL & NLO & LO \\\hline\hline
0.5 
 & $4006^{+232(5.8\%)}_{-276(6.9\%)}~^{+466(11.7\%)}_{-332(8.3\%)}$ 
 & $3802^{+342(9\%)}_{-421(11.1\%)}~^{+455(12.0\%)}_{-322(8.5\%)}$
 & $2726^{+876(32.1\%)}_{-618(22.7\%)}~^{+314(11.5\%)}_{-221(8.1\%)}$
\\\hline
0.6 
 & $1429^{+76.3(5.3\%)}_{-93.2(6.5\%)}~^{+195(13.7\%)}_{-134(9.4\%)}$
 & $1352^{+116(8.6\%)}_{-148(11.0\%)}~^{+188(14.0\%)}_{-129(9.5\%)}$
 & $980.7^{+319(32.5\%)}_{-225(22.9\%)}~^{+130(13.3\%)}_{-88.7(9.1\%)}$
\\\hline
0.7 
 & $577.6^{+28.7(5\%)}_{-36.0(6.2\%)}~^{+89.0(15.4\%)}_{-59.6(10.3\%)}$
 & $545.1^{+44.9(8.2\%)}_{-59.3(10.9\%)}~^{+85.5(15.7\%)}_{-57.0(10.5\%)}$
 & $399.1^{+131(32.9\%)}_{-92.1(23.1\%)}~^{+59.0(14.8\%)}_{-39.2(9.8\%)}$
\\\hline
0.8 
 & $256.0^{+12.0(4.7\%)}_{-15.4(6\%)}~^{+43.4(17\%)}_{-28.4(11.1\%)}$
 & $241.0^{+19.3(8\%)}_{-26.1(10.8\%)}~^{+41.4(17.2\%)}_{-27.0(11.2\%)}$
 & $177.8^{+58.9(33.1\%)}_{-41.3(23.2\%)}~^{+28.5(16.1\%)}_{-18.6(10.5\%)}$
\\\hline
0.9 
 & $121.7^{+5.41(4.4\%)}_{-7.06(5.8\%)}~^{+22.3(18.4\%)}_{-14.4(11.8\%)}$
 & $114.3^{+8.95(7.8\%)}_{-12.3(10.8\%)}~^{+21.2(18.5\%)}_{-13.5(11.9\%)}$
 & $84.71^{+28.3(33.4\%)}_{-19.8(23.4\%)}~^{+14.6(17.2\%)}_{-9.32(11\%)}$
\\\hline
1.0 
 & $61.12^{+2.59(4.2\%)}_{-3.45(5.6\%)}~^{+12.0(19.6\%)}_{-7.58(12.4\%)}$
 & $57.25^{+4.42(7.7\%)}_{-6.17(10.8\%)}~^{+11.3(19.7\%)}_{-7.11(12.4\%)}$
 & $42.57^{+14.3(33.7\%)}_{-10.0(23.5\%)}~^{+7.75(18.2\%)}_{-4.88(11.5\%)}$
\\\hline
1.1 
 & $32.05^{+1.32(4.1\%)}_{-1.76(5.5\%)}~^{+6.68(20.8\%)}_{-4.15(13\%)}$
 & $29.94^{+2.29(7.7\%)}_{-3.24(10.8\%)}~^{+6.25(20.9\%)}_{-3.88(12.9\%)}$
 & $22.31^{+7.56(33.9\%)}_{-5.28(23.6\%)}~^{+4.28(19.2\%)}_{-2.66(11.9\%)}$
\\\hline
1.2 
 & $17.41^{+0.706(4.1\%)}_{-0.939(5.4\%)}~^{+3.83(22\%)}_{-2.35(13.5\%)}$
 & $16.23^{+1.24(7.6\%)}_{-1.76(10.9\%)}~^{+3.57(22\%)}_{-2.18(13.4\%)}$
 & $12.10^{+4.13(34.2\%)}_{-2.88(23.8\%)}~^{+2.43(20.1\%)}_{-1.50(12.4\%)}$
\\\hline
1.3 
 & $9.737^{+0.388(4\%)}_{-0.516(5.3\%)}~^{+2.25(23.2\%)}_{-1.36(14\%)}$
 & $9.049^{+0.693(7.7\%)}_{-0.989(10.9\%)}~^{+2.09(23.1\%)}_{-1.26(13.9\%)}$
 & $6.745^{+2.32(34.4\%)}_{-1.61(23.9\%)}~^{+1.42(21.1\%)}_{-0.864(12.8\%)}$
\\\hline
1.4 
 & $5.578^{+0.218(3.9\%)}_{-0.291(5.2\%)}~^{+1.36(24.3\%)}_{-0.810(14.5\%)}$
 & $5.169^{+0.398(7.7\%)}_{-0.569(11\%)}~^{+1.25(24.2\%)}_{-0.745(14.4\%)}$
 & $3.848^{+1.34(34.7\%)}_{-0.927(24.1\%)}~^{+0.850(22.1\%)}_{-0.511(13.3\%)}$
\\\hline
1.5
 & $3.260^{+0.126(3.9\%)}_{-0.168(5.2\%)}~^{+0.833(25.5\%)}_{-0.492(15.1\%)}$
 & $3.012^{+0.235(7.8\%)}_{-0.335(11.1\%)}~^{+0.763(25.3\%)}_{-0.450(14.9\%)}$
 & $2.238^{+0.783(35\%)}_{-0.543(24.2\%)}~^{+0.518(23.1\%)}_{-0.309(13.8\%)}$
\\\hline
1.6 
 & $1.938^{+0.074(3.8\%)}_{-0.099(5.1\%)}~^{+0.520(26.8\%)}_{-0.304(15.7\%)}$
 & $1.785^{+0.141(7.9\%)}_{-0.200(11.2\%)}~^{+0.474(26.5\%)}_{-0.277(15.5\%)}$
 & $1.323^{+0.467(35.3\%)}_{-0.323(24.4\%)}~^{+0.321(24.3\%)}_{-0.190(14.4\%)}$
\\\hline
1.7 
 & $1.169^{+0.044(3.7\%)}_{-0.059(5.1\%)}~^{+0.329(28.2\%)}_{-0.191(16.3\%)}$
 & $1.073^{+0.086(8\%)}_{-0.122(11.4\%)}~^{+0.299(27.8\%)}_{-0.173(16.1\%)}$
 & $0.793^{+0.282(35.6\%)}_{-0.195(24.6\%)}~^{+0.202(25.5\%)}_{-0.119(15\%)}$
\\\hline
1.8 
 & $0.714^{+0.026(3.7\%)}_{-0.036(5\%)}~^{+0.212(29.6\%)}_{-0.123(17.2\%)}$
 & $0.653^{+0.053(8.2\%)}_{-0.075(11.5\%)}~^{+0.191(29.2\%)}_{-0.109(16.8\%)}$
 & $0.480^{+0.173(35.9\%)}_{-0.119(24.7\%)}~^{+0.129(26.9\%)}_{-0.075(15.7\%)}$
\\\hline
1.9 
 & $0.440^{+0.016(3.7\%)}_{-0.022(5\%)}~^{+0.137(31.2\%)}_{-0.078(17.7\%)}$
 & $0.401^{+0.033(8.4\%)}_{-0.047(11.7\%)}~^{+0.123(30.8\%)}_{-0.070(17.5\%)}$
 & $0.294^{+0.107(36.3\%)}_{-0.073(24.9\%)}~^{+0.083(28.3\%)}_{-0.048(16.4\%)}$
\\\hline
2.0 
 & $0.274^{+0.010(3.6\%)}_{-0.013(5\%)}~^{+0.090(32.9\%)}_{-0.051(18.6\%)}$
 & $0.248^{+0.021(8.5\%)}_{-0.029(11.8\%)}~^{+0.080(32.4\%)}_{-0.045(18.3\%)}$
 & $0.181^{+0.066(36.6\%)}_{-0.045(25.1\%)}~^{+0.054(30\%)}_{-0.031(17.3\%)}$
\\\hline
\end{tabular}
\caption{\label{tab:HF_CTEQ65}
Cross sections (in $fb$) for the production of $T\bar{T}$ pairs at the LHC,
computed with CTEQ6.5 PDFs. The mass of the heavy fermion $T$ is expressed 
in TeV. For each entry of the table, we give the central value of the cross 
section, with the scale and PDF uncertainties.}
\end{center}
\end{table}

\begin{table}
\begin{center}
\begin{tabular}{|c|c|c|c|}
\hline
$m_T$ & NLO+NLL & NLO & LO \\\hline\hline
0.5 
 & $4462^{+267.4(6\%)}_{-314.0(7\%)}~^{+197.4(4.4\%)}_{-172.6(3.9\%)}$ 
 & $4236^{+392.9(9.3\%)}_{-480.5(11.3\%)}~^{+191.9(4.5\%)}_{-167.7(4\%)}$ 
 & $3017^{+988.3(32.8\%)}_{-694.6(23\%)}~^{+132.0(4.4\%)}_{-115.9(3.8\%)}$ 
\\\hline
0.6 
 & $1599^{+88.7(5.5\%)}_{-107.0(6.7\%)}~^{+81.2(5.1\%)}_{-70.1(4.4\%)}$ 
 & $1513^{+134.9(8.9\%)}_{-170.9(11.3\%)}~^{+78.2(5.2\%)}_{-67.5(4.5\%)}$ 
 & $1089^{+362.9(33.3\%)}_{-253.8(23.3\%)}~^{+53.8(4.9\%)}_{-46.6(4.3\%)}$ 
\\\hline
0.7 
 & $648.8^{+33.7(5.2\%)}_{-41.6(6.4\%)}~^{+36.3(5.6\%)}_{-31.0(4.8\%)}$ 
 & $611.9^{+52.9(8.6\%)}_{-68.9(11.3\%)}~^{+34.7(5.7\%)}_{-29.6(4.8\%)}$ 
 & $444.0^{+145.0(33.8\%)}_{-104.5(23.5\%)}~^{+23.9(5.4\%)}_{-20.4(4.6\%)}$ 
\\\hline
0.8 
 & $288.2^{+14.1(4.9\%)}_{-17.8(6.2\%)}~^{+17.3(6\%)}_{-14.6(5.1\%)}$ 
 & $271.0^{+22.9(8.4\%)}_{-30.5(11.2\%)}~^{+16.4(6.1\%)}_{-13.9(5.1\%)}$ 
 & $198.0^{+67.5(34.1\%)}_{-46.9(23.7\%)}~^{+11.4(5.7\%)}_{-9.57(4.8\%)}$ 
\\\hline
0.9 
 & $137.2^{+6.42(4.7\%)}_{-8.21(6\%)}~^{+8.72(6.4\%)}_{-7.28(5.3\%)}$ 
 & $128.6^{+10.7(8.3\%)}_{-14.5(11.3\%)}~^{+8.23(6.4\%)}_{-6.85(5.3\%)}$ 
 & $94.34^{+32.5(34.4\%)}_{-22.5(23.9\%)}~^{+5.74(6.1\%)}_{-4.72(5\%)}$ 
\\\hline
1.0 
 & $68.97^{+3.09(4.5\%)}_{-4.02(5.8\%)}~^{+4.61(6.7\%)}_{-3.77(5.5\%)}$ 
 & $64.48^{+5.30(8.2\%)}_{-7.27(11.3\%)}~^{+4.32(6.7\%)}_{-3.52(5.5\%)}$ 
 & $47.43^{+16.5(34.8\%)}_{-11.4(24\%)}~^{+3.05(6.4\%)}_{-2.44(5.1\%)}$ 
\\\hline
1.1 
 & $36.21^{+1.56(4.3\%)}_{-2.06(5.7\%)}~^{+2.53(7\%)}_{-2.03(5.6\%)}$ 
 & $33.75^{+2.76(8.2\%)}_{-3.82(11.3\%)}~^{+2.37(7\%)}_{-1.88(5.6\%)}$ 
 & $24.87^{+8.72(35\%)}_{-6.02(24.2\%)}~^{+1.69(6.8\%)}_{-1.31(5.3\%)}$ 
\\\hline
1.2 
 & $19.69^{+0.824(4.2\%)}_{-1.10(5.6\%)}~^{+1.44(7.3\%)}_{-1.12(5.7\%)}$ 
 & $18.30^{+1.49(8.2\%)}_{-2.08(11.4\%)}~^{+1.35(7.4\%)}_{-1.04(5.7\%)}$ 
 & $13.50^{+4.77(35.3\%)}_{-3.29(24.4\%)}~^{+0.987(7.3\%)}_{-0.734(5.4\%)}$ 
\\\hline
1.3 
 & $11.03^{+0.449(4.1\%)}_{-0.604(5.5\%)}~^{+0.857(7.8\%)}_{-0.642(5.8\%)}$ 
 & $10.22^{+0.837(8.2\%)}_{-1.17(11.5\%)}~^{+0.802(7.9\%)}_{-0.593(5.8\%)}$ 
 & $7.533^{+2.68(35.6\%)}_{-1.85(24.5\%)}~^{+0.597(7.9\%)}_{-0.425(5.6\%)}$ 
\\\hline
1.4 
 & $6.324^{+0.252(4\%)}_{-0.341(5.4\%)}~^{+0.524(8.3\%)}_{-0.377(6\%)}$ 
 & $5.843^{+0.482(8.3\%)}_{-0.674(11.5\%)}~^{+0.492(8.4\%)}_{-0.348(6\%)}$ 
 & $4.303^{+1.55(35.9\%)}_{-1.06(24.7\%)}~^{+0.372(8.7\%)}_{-0.253(5.9\%)}$ 
\\\hline
1.5
 & $3.702^{+0.145(3.9\%)}_{-0.197(5.3\%)}~^{+0.329(8.9\%)}_{-0.227(6.1\%)}$ 
 & $3.409^{+0.284(8.3\%)}_{-0.397(11.6\%)}~^{+0.310(9.1\%)}_{-0.210(6.2\%)}$ 
 & $2.506^{+0.908(36.2\%)}_{-0.622(24.8\%)}~^{+0.238(9.5\%)}_{-0.155(6.2\%)}$ 
\\\hline
1.6 
 & $2.203^{+0.085(3.8\%)}_{-0.116(5.3\%)}~^{+0.211(9.6\%)}_{-0.140(6.3\%)}$ 
 & $2.022^{+0.171(8.5\%)}_{-0.238(11.8\%)}~^{+0.199(9.9\%)}_{-0.130(6.4\%)}$ 
 & $1.483^{+0.542(36.5\%)}_{-0.370(25\%)}~^{+0.155(10.4\%)}_{-0.097(6.5\%)}$ 
\\\hline
1.7 
 & $1.330^{+0.051(3.8\%)}_{-0.069(5.2\%)}~^{+0.138(10.4\%)}_{-0.088(6.6\%)}$ 
 & $1.217^{+0.104(8.6\%)}_{-0.145(11.9\%)}~^{+0.131(10.7\%)}_{-0.082(6.7\%)}$ 
 & $0.890^{+0.328(36.8\%)}_{-0.224(25.1\%)}~^{+0.102(11.5\%)}_{-0.061(6.9\%)}$ 
\\\hline
1.8 
 & $0.813^{+0.031(3.8\%)}_{-0.042(5.2\%)}~^{+0.091(11.2\%)}_{-0.056(6.9\%)}$ 
 & $0.741^{+0.065(8.8\%)}_{-0.089(12.1\%)}~^{+0.087(11.7\%)}_{-0.052(7\%)}$ 
 & $0.540^{+0.201(37.2\%)}_{-0.137(25.3\%)}~^{+0.068(12.6\%)}_{-0.039(7.3\%)}$ 
\\\hline
1.9 
 & $0.502^{+0.019(3.7\%)}_{-0.026(5.1\%)}~^{+0.061(12.2\%)}_{-0.036(7.2\%)}$ 
 & $0.455^{+0.041(8.9\%)}_{-0.056(12.2\%)}~^{+0.058(12.8\%)}_{-0.034(7.4\%)}$ 
 & $0.330^{+0.124(37.5\%)}_{-0.084(25.5\%)}~^{+0.046(13.8\%)}_{-0.026(7.8\%)}$ 
\\\hline
2.0 
 & $0.312^{+0.012(3.7\%)}_{-0.016(5.1\%)}~^{+0.041(13.2\%)}_{-0.023(7.5\%)}$ 
 & $0.282^{+0.026(9.1\%)}_{-0.035(12.4\%)}~^{+0.039(13.9\%)}_{-0.022(7.8\%)}$ 
 & $0.204^{+0.077(37.9\%)}_{-0.052(25.7\%)}~^{+0.031(15\%)}_{-0.017(8.2\%)}$ 
\\\hline
\end{tabular}
\caption{\label{tab:HF_MRST}
Cross sections (in $fb$) for the production of $T\bar{T}$ pairs at the LHC,
computed with MRST2006nnlo PDFs. The mass of the heavy fermion $T$ is 
expressed in TeV. For each entry of the table, we give the central value 
of the cross section, with the scale and PDF uncertainties.}
\end{center}
\end{table}
Our results are presented in tables~\ref{tab:HF_CTEQ65} and~~\ref{tab:HF_MRST}.
A few comments are in order.
\begin{itemize}
\item The scale dependence of the NLO+NLL cross section is small for
all values of $m_T$, and decreases monotonically with increasing $m_T$.
This is to be expected, since the heavier the fermion, the closer the
kinematics is to the threshold.
\item The scale dependence of the NLO cross section starts by decreasing,
but then tends to increase with increasing $m_T$. This is the signal of
the necessity of including threshold corrections. On the other hand,
the scale dependence of the LO cross section is always extremely large.
This fact must be taken into proper account if an estimate of
the $K$ factor is needed. In particular, one must precisely understand
which hard scale is used in a LO computation (e.g. in a standard
parton-shower Monte Carlo).
\item The relative PDF uncertainty is extremely large in the case
of CTEQ6.5. When MRST2006nnlo sets are used, that uncertainty is smaller
by a factor of about 2--3, consistently with what already observed
in the case of top production. By and large, the PDF uncertainty affects 
equally the NLO+NLL, the NLO, and the LO cross sections. At the largest
$m_T$ values considered here, it prevents one from giving a precise
prediction even in the case of the NLO+NLL computation. It will 
therefore be highly desirable to measure the PDFs at the LHC for
intermediate- and large-$x$ regions, using e.g. 
low-mass final states produced at large rapidity.
\end{itemize}

\section{Conclusions}

In this paper we have produced and tabulated updated predictions for the
next-to-leading order plus next-to-leading log resummed (NLO+NLL) cross
sections for $t\bar t$ production at the Tevatron and at the LHC. 
QCD cross sections for heavy fermion production at the LHC are also given.

The theoretical uncertainties due to unknown higher orders and to the imperfect
knowledge of the parton
distribution function sets are explored in detail
and also tabulated. NLO and LO results are also given, for reference and
comparison.

The main results for $t\bar t$ production at the LHC ($\sqrt{S} = 14$~TeV) 
are the following:
\begin{equation}
\sigma^\mathrm{NLO+NLL}_{t\bar t}(\mathrm{LHC}, m_t = 171~\mathrm{GeV},
\mathrm{CTEQ6.5})
= 908 {~}^{+82(9.0\%)}_{-85(9.3\%)}~\mathrm{(scales)} 
{~}^{+30(3.3\%)}_{-29(3.2\%)} ~\mathrm{(PDFs)} ~~\mathrm{pb}
\end{equation}
\begin{equation}
\sigma^\mathrm{NLO+NLL}_{t\bar t}(\mathrm{LHC}, m_t = 171~\mathrm{GeV},
\mathrm{MRST2006nnlo})
= 961 {~}^{+89(9.2\%)}_{-91(9.4\%)}~\mathrm{(scales)} 
{~}^{+11(1.1\%)}_{-12(1.2\%)} ~\mathrm{(PDFs)} ~~\mathrm{pb}
\end{equation}
Cross sections obtained with two of the most recent PDF sets are
given, since they appear to be only partially compatible within their respective
uncertainties.

Finally, we note that ref.~\cite{Moch:2008qy} recently produced an approximated
NNLO cross section by truncating a soft-gluon NNLL resummed calculation to
order $\alpha_s^4$. Their phenomenological analysis produces cross sections for
the LHC with extremely small scales uncertainty, of order 2-3\%, sensibly
smaller than ours. 
We have argued in section.~\ref{sec:disc} that such a small uncertainty also arises at the NLO+NLL level by requiring the factorization and renormalization scales to be equal. It will therefore be interesting to verify whether such reduced scale dependence found in~\cite{Moch:2008qy}  survives a test with independent scales, thus showing a genuine improvement due to the added NNLO terms, or whether it is an intrinsic consequence of keeping the scales equal.

\section*{Note added}
After this work was completed, a new study of the $t\bar{t}$ cross section at the Tevatron and LHC appeared in ref.~\cite{Kidonakis:2008mu}.

\section*{Acknowledgements}
We thank R.~Contino for discussions on heavy fermion pair production.
This work is supported in part by the European Community's 
Marie-Curie Research Training Network HEPTOOLS under contract 
MRTN-CT-2006-035505. MC is supported in part by grant ANR-05-JCJC-0046-01 from 
the French Agence Nationale de la Recherche.

\appendix
\section{On scale and PDF uncertainties\label{sec:details}}
The definitions we have adopted in eqs.~(\ref{eq:muplus}) 
and~(\ref{eq:muminus}) for scale of uncertainty, and in 
eqs.~(\ref{eq:PDFplus}) and~(\ref{eq:PDFminus})
for PDF uncertainty, are by no means unique. In this
section, we shall briefly illustrate other choices.

The possibility of making different choices stems from the observation
that scale and PDF uncertainties have to be combined in order to obtain
an estimate of the overall uncertainty affecting the cross section.
The way in which this combination is to be performed is at present unclear,
given the fact that neither the scale uncertainty nor the PDF uncertainty
(the latter owing to the fact that PDF error sets are derived in violation
of the $\Delta\chi^2=1$ rule) follow the laws of statistical errors.

Scale uncertainty can in general be written as
\begin{eqnarray}
&&\Delta{\sigma}_{\mu +} = \sigma\left(\xiFM,\xiRM\right)-\sigma(1,1)\,,
\label{eq:muplus2}
\\
&&\Delta{\sigma}_{\mu -} = \sigma(1,1)-\sigma\left(\xiFm,\xiRm\right)\,,
\label{eq:muminus2}
\end{eqnarray}
where different prescriptions can be devised for the determination
of $(\xiFM,\xiRM)$ and of\\ $(\xiFm,\xiRm)$. 
As far as PDF uncertainty is concerned, one always makes use of 
\begin{eqnarray}
&&\delta{\sigma}_{\sss PDF+}(\xiF,\xiR) = 
\sqrt{\sum_i\Big(\max\Big[{\sigma}(set_{+i}) - {\sigma}(set_0),
 {\sigma}(set_{-i}) - {\sigma}(set_0), 0\Big]\Big)^2} \,, 
\label{eq:PDFplus2}
\\
&&\delta{\sigma}_{\sss PDF-}(\xiF,\xiR) = 
\sqrt{\sum_i\Big(\max\Big[{\sigma}(set_0) - {\sigma}(set_{+i}),
 {\sigma}(set_0) - {\sigma}(set_{-i}), 0\Big]\Big)^2} \,,
\label{eq:PDFminus2}
\end{eqnarray}
and then defines
\begin{eqnarray}
&&\Delta{\sigma}_{\sss PDF+} = 
\delta{\sigma}_{\sss PDF+}\left(\xiFMb,\xiRMb\right)\,,
\label{eq:PDFplus3}
\\
&&\Delta{\sigma}_{\sss PDF-} = 
\delta{\sigma}_{\sss PDF-}\left(\xiFmb,\xiRmb\right)\,,
\label{eq:PDFminus3}
\end{eqnarray}
where again the values of $(\xiFMb,\xiRMb)$ and of $(\xiFmb,\xiRmb)$ at 
which the r.h.s. of these equations are evaluated are a matter of choice.

We limit ourselves to give four examples.
\begin{itemize}
\item[{\em A)}] Our default choice, illustrated in sect.~\ref{sec:top}
and which gives rise to the results presented in this paper, is
equivalent to solving
\begin{eqnarray}
\max_{\{\xiF,\xiR\}}\Big[\sigma(\xiF,\xiR)-\sigma(1,1)\Big]&=&
\sigma\left(\xiFM,\xiRM\right)-\sigma(1,1)\,,
\label{eq:muplusA}
\\
\min_{\{\xiF,\xiR\}}\Big[\sigma(\xiF,\xiR)-\sigma(1,1)\Big]&=&
\sigma\left(\xiFm,\xiRm\right)-\sigma(1,1)\,,
\label{eq:muminusA}
\end{eqnarray}
for $(\xiFM,\xiRM)$ and $(\xiFm,\xiRm)$, which are then used 
in eqs.~(\ref{eq:muplus2}) and~(\ref{eq:muminus2}). The PDF uncertainty
is defined by setting
\begin{eqnarray}
\left(\xiFMb,\xiRMb\right)=(1,1)\,,\;\;\;\;\;\;
\left(\xiFmb,\xiRmb\right)=(1,1)\,.
\end{eqnarray}
\item[{\em B)}] The scale uncertainty is defined in the same way as done
in item {\em A)}. For the PDF uncertainty, we set
\begin{eqnarray}
\left(\xiFMb,\xiRMb\right)=\left(\xiFM,\xiRM\right)\,,\;\;\;\;\;\;
\left(\xiFmb,\xiRmb\right)=\left(\xiFm,\xiRm\right)\,.
\end{eqnarray}
with $(\xiFM,\xiRM)$ and $(\xiFm,\xiRm)$ computed again as in 
eqs.~(\ref{eq:muplusA}) and~(\ref{eq:muminusA}).
\item[{\em C)}] We first solve
\begin{eqnarray}
&&\max_{\{\xiF,\xiR\}}\Big[\sigma(\xiF,\xiR)+
\delta{\sigma}_{\sss PDF+}(\xiF,\xiR)-\sigma(1,1)\Big]=
\nonumber\\*&&\phantom{aaaaaaaa}
\sigma\left(\xiFM,\xiRM\right)+
\delta{\sigma}_{\sss PDF+}\left(\xiFM,\xiRM\right)-\sigma(1,1)\,,
\label{eq:muplusC}
\\
&&\min_{\{\xiF,\xiR\}}\Big[\sigma(\xiF,\xiR)-
\delta{\sigma}_{\sss PDF-}(\xiF,\xiR)-\sigma(1,1)\Big]=
\nonumber\\*&&\phantom{aaaaaaaa}
\sigma\left(\xiFm,\xiRm\right)-
\delta{\sigma}_{\sss PDF-}\left(\xiFm,\xiRm\right)-\sigma(1,1)\,,
\label{eq:muminusC}
\end{eqnarray}
for $(\xiFM,\xiRM)$ and $(\xiFm,\xiRm)$. These values are then used
to determine the scale uncertainty according to eqs.~(\ref{eq:muplus}) 
and~(\ref{eq:muminus}), and the PDF uncertainty by setting
\begin{eqnarray}
\left(\xiFMb,\xiRMb\right)=\left(\xiFM,\xiRM\right)\,,\;\;\;\;\;\;
\left(\xiFmb,\xiRmb\right)=\left(\xiFm,\xiRm\right)\,,
\end{eqnarray}
which are then used in eqs~(\ref{eq:PDFplus3}) and~(\ref{eq:PDFminus3}).
\item[{\em D)}] The scale uncertainty is defined in the same way as done
in item {\em A)}. For PDF uncertainty, we solve
\begin{eqnarray}
\max_{\{\xiF,\xiR\}}\Big[\delta{\sigma}_{\sss PDF+}(\xiF,\xiR)\Big]&=&
\delta{\sigma}_{\sss PDF+}\left(\xiFMb,\xiRMb\right)\,,
\label{eq:PDFplusD}
\\
\max_{\{\xiF,\xiR\}}\Big[\delta{\sigma}_{\sss PDF-}(\xiF,\xiR)\Big]&=&
\delta{\sigma}_{\sss PDF-}\left(\xiFmb,\xiRmb\right)\,,
\label{eq:PDFminusD}
\end{eqnarray}
for $(\xiFMb,\xiRMb)$ and $(\xiFmb,\xiRmb)$, which are then used 
in eqs~(\ref{eq:PDFplus3}) and~(\ref{eq:PDFminus3}).
\end{itemize}

Items {\em B)}-{\em D)} follow the same logic, namely finding
the absolute maximum and minimum of the cross section, by various
combinations of scale and PDF uncertainties. In this sense, it is not
fully justified to quote these two uncertainties separately, although 
it is still convenient for bookkeeping. These approaches stem from
the observation that, in a hadroproduction QCD computation, unknown
higher orders also enter the determination of the PDFs, and one is
therefore entitled to use the full information on the PDF uncertainty
in the determination of the scale dependence. In fact, the three
methods give very similar results, with {\em D)} being the most
conservative, i.e. resulting in the largest overall cross section
uncertainty.

On the other hand, by following the procedure outlined in item {\em A)},
one is able to better assess the separate dependence upon scales and PDFs.
It should be observed that, while the quantities
$\delta{\sigma}_{\sss PDF\pm}(\xiF,\xiR)$ depend on $\xiF$ and $\xiR$
roughly in the same way as the cross sections $\sigma(\xiF,\xiR)$,
the ratios
\begin{equation}
\delta{\sigma}_{\sss PDF\pm}(\xiF,\xiR)\Big/
\sigma(\xiF,\xiR)
\label{PDFrelerr}
\end{equation}
are extremely stable with respect to variations of $\xiF$ and $\xiR$.
This implies that the {\em relative} PDF uncertainty on the central
value of the cross section,
that is
\begin{equation}
\Delta{\sigma}_{\sss PDF\pm}\Big/\sigma(1,1)
\end{equation}
is basically identical to any of those in eq.~(\ref{PDFrelerr}), if one
follows item {\em A)}. This is not the case for items {\em B)}--{\em D)};
the relative uncertainty due to $\Delta{\sigma}_{\sss PDF+}$
($\Delta{\sigma}_{\sss PDF-}$) tends to be larger (smaller) than
that computed according to item {\em A)}.

The consideration above led us to prefer the procedure of item {\em A)}
for the computation of the results presented in this paper. This has
also the advantage that it renders the calculation less demanding from
the point of view of CPU time. We conclude by stressing that for top
production the procedures of items {\em B)}--{\em D)} would have given 
similar results as that of item {\em A)}.

\end{document}